\documentclass[acmsmall]{acmart}

\setcopyright{acmlicensed}
\acmDOI{10.1145/3660783}
\acmYear{2024}
\copyrightyear{2024}
\acmSubmissionID{fse24main-p220-p}
\acmJournal{PACMSE}
\acmVolume{1}
\acmNumber{FSE}
\acmArticle{76}
\acmMonth{7}
\received{2023-09-28}
\received[accepted]{2024-04-16}


\usepackage{adjustbox}
\usepackage{amsmath,amssymb,amsfonts}
\usepackage[linesnumbered,ruled,vlined]{algorithm2e}
\usepackage{setspace}
\usepackage{algorithmic}
\usepackage{float}
\usepackage{graphicx}
\usepackage{textcomp}
\usepackage{xcolor}
\usepackage{multirow}
\usepackage{mdframed}
\usepackage{balance}
\usepackage{booktabs}
\usepackage{tcolorbox}
\usepackage{makecell}
\usepackage{threeparttable}
\usepackage{colortbl}
\usepackage{subfigure}
\usepackage{hhline}
\usepackage{pifont}
\usepackage{listings}
\usepackage{enumitem}

\newcommand{\tabincell}[2]{\begin{tabular}{@{}#1@{}}#2\end{tabular}}  

\newcommand{\parabf}[1]{\noindent\textbf{#1}}

\newcommand{\revised}[1]{\textcolor{black}{\textbf{}#1}}

\definecolor{ggray}{HTML}{eff0f0}
\definecolor{gggray}{HTML}{E8E8E8}
\definecolor{ggggray}{HTML}{BEBEBE}

\newcommand{\ie}{\textit{i.e.,}}
\newcommand{\eg}{\textit{e.g.,}}

\newcommand{\chatgpt}{ChatGPT}
\newcommand{\athenatest}{AthenaTest}
\newcommand{\evosuite}{Evosuite}

\newcommand{\codellama}{CodeLlama-Instruct}
\newcommand{\codefuse}{CodeFuse}

\newcommand{\ourtool}{\textsc{ChatTester}}
\newcommand{\ourtoolm}{\textsc{ChatTester-Ite}}
\newcommand{\ourtooIte}{\textsc{ChatTester-Ini}}

\usepackage[utf8]{inputenc}
\usepackage[english]{babel}

\newcounter{finding}
\newcommand{\finding}[1]{\refstepcounter{finding}
 	\vspace{1mm}
	\begin{mdframed}[linecolor=gray,roundcorner=12pt,backgroundcolor=gray!15,linewidth=3pt,innerleftmargin=2pt, leftmargin=0cm,rightmargin=0cm,topline=false,bottomline=false,rightline = false]
		\textbf{Finding \arabic{finding}:} #1
	\end{mdframed}
	\vspace{3mm}
}

\newcommand{\distance}{5pt}
\begin{abstract}
Unit testing plays an essential role in detecting bugs in functionally-discrete program units (\eg{} methods). Manually writing high-quality unit tests is time-consuming and laborious. Although the traditional techniques are able to generate tests with reasonable coverage, they are shown to exhibit low readability and still cannot be directly adopted by developers in practice. Recent work has shown the large potential of large language models (LLMs) in unit test generation. By being pre-trained on a massive developer-written code corpus, the models are capable of generating more human-like and meaningful test code. 

In this work, we perform the first empirical study to evaluate the capability of \chatgpt{} (\ie{} one of the most representative LLMs with outstanding performance in code generation and comprehension) in unit test generation. In particular, we conduct both a quantitative analysis and a user study to systematically investigate the quality of its generated tests in terms of correctness, sufficiency, readability, and usability. We find that the tests generated by \chatgpt{} still suffer from correctness issues, including diverse compilation errors and execution failures (mostly caused by incorrect assertions); but the passing tests generated by \chatgpt{} almost resemble manually-written tests by achieving comparable coverage, readability, and even sometimes developers' preference. Our findings indicate that generating unit tests with \chatgpt{} could be very promising if the correctness of its generated tests could be further improved. 

Inspired by our findings above, we further propose \ourtool{}, a novel \chatgpt -based unit test generation approach, which leverages \chatgpt{} itself to improve the quality of its generated tests. \ourtool{} incorporates an initial test generator and an iterative test refiner. Our evaluation demonstrates the effectiveness of \ourtool{} by generating 34.3\% more compilable tests and 18.7\% more tests with correct assertions than the default \chatgpt{}. 
In addition to \chatgpt{}, we further investigate the generalization capabilities of \ourtool{} by applying it to two recent open-source LLMs (\ie{} \codellama{} and \codefuse{}) and our results show that  \ourtool{} can also improve the quality of tests generated by these LLMs.

\end{abstract}


\begin{document}

\title{No More Manual Tests? Evaluating and Improving ChatGPT for Unit Test Generation}

\author{Zhiqiang Yuan}
\orcid{0000-0002-6497-9380}
\affiliation{%
  \institution{Department of Computer Science, Fudan University}
  \city{Shanghai}
  \country{China}
}
\email{zhiqiangyuan23@m.fudan.edu.cn}

\author{Mingwei Liu}
\orcid{0000-0002-3462-997X}
\affiliation{%
  \institution{Department of Computer Science, Fudan University}
  \city{Shanghai}
  \country{China}
}
\email{liumingwei@fudan.edu.cn}

\author{Shiji Ding}
\orcid{0009-0007-7781-4095}
\affiliation{%
  \institution{Department of Computer Science, Fudan University}
  \city{Shanghai}
  \country{China}
}
\email{sjding22@m.fudan.edu.cn}

\author{Kaixin Wang}
\orcid{0009-0002-6414-2174}
\affiliation{%
  \institution{Department of Computer Science, Fudan University}
  \city{Shanghai}
  \country{China}
}
\email{kxwang23@m.fudan.edu.cn}

\author{Yixuan Chen}
\orcid{0009-0006-8804-1828}
\affiliation{%
  \institution{Department of Computer Science, Fudan University}
  \city{Shanghai}
  \country{China}
}
\email{yixuanchen23@m.fudan.edu.cn}

\author{Xin Peng}
\orcid{0000-0003-3376-2581}
\affiliation{%
  \institution{Department of Computer Science, Fudan University}
  \city{Shanghai}
  \country{China}
}
\email{pengxin@fudan.edu.cn}

\author{Yiling Lou$^\ast$}
\orcid{0000-0002-4066-3365}
\affiliation{%
  \institution{Department of Computer Science, Fudan University}
  \city{Shanghai}
  \country{China}
  \authornote{Corresponding author (yilinglou@fudan.edu.cn)}
}
\email{yilinglou@fudan.edu.cn}

\renewcommand{\shortauthors}{Yuan, et al.}

\maketitle

\section{Introduction}

Unit testing~\cite{zhu1997software, runeson2006survey, olan2003unit} validates whether a functionally-discrete program unit (\eg{} a method) under test behaves correctly. 
As the primary stage in the software development procedure, unit testing plays an essential role in detecting and diagnosing bugs in a nascent stage and prevents their further propagation in the development cycle. 
Therefore, writing high-quality unit tests is crucial for ensuring software quality.
For a method under test (\ie{} often called as the focal method), its corresponding unit test consists of a \textit{test prefix} and a \textit{test oracle}~\cite{DBLP:journals/tse/BarrHMSY15}.  
In particular, the test prefix is typically a series of method invocation statements or assignment statements, which aim at driving the focal method to a testable state; and then the test oracle serves as the specification to check whether the current behavior of the focal method satisfies the expected one. 

Manually writing and maintaining high-quality unit tests can be very time-consuming and laborious~\cite{DBLP:conf/icst/KlammerK15, DBLP:conf/issre/DakaF14}. To alleviate manual efforts in writing unit tests, researchers have proposed various techniques to facilitate automated test generation~\cite{DBLP:journals/ese/LukasczykKF23, DBLP:conf/icse/LukasczykF22, DBLP:journals/jss/KurianBBD23, DBLP:conf/kbse/ScalabrinoGNGLG18, DBLP:conf/kbse/BlasiGEP22}. 
Traditional unit test generation techniques leverage search-based~\cite{harman2009theoretical, blasi2022call, delgado2022interevo}, constraint-based~\cite{ernst2007daikon, csallner2008dysy, xiao2013characteristic}, or random-based strategies~\cite{zeller2019fuzzing, pacheco2007feedback} to generate a suite of unit tests with the main goal of maximizing the coverage in the software under test. 
Although achieving reasonable coverage, these automatically-generated tests exhibit a large gap to manual-written ones in terms of readability and meaningfulness, and thus developers are mostly unwilling to directly adopt them in practice ~\cite{almasi2017industrial}. 

To address these issues, recent work~\cite{tufano2020unit, nie2023learning, watson2020learning, dinella2022toga, mastropaolo2021studying} has leveraged advanced deep learning (DL) techniques, especially large language models (LLMs), to generate unit tests. 
These techniques mostly formulate unit test generation as a neural machine translation problem by translating a given focal method into the corresponding test prefix and the test assertion. 
In particular, they incorporate the power of LLMs by fine-tuning these pre-trained models on the test generation task. 
Owing to being extensively pre-trained on a massive developer-written code corpus and then being specifically fine-tuned on the test generation task, these models are capable of generating more human-like and meaningful test code, showing a large potential of LLMs in unit test generation.

\parabf{Study.} In this work, we perform the first empirical study to evaluate the capability of \chatgpt~\cite{chatGPT} (\ie{} one of the most representative LLMs) in unit test generation. Different from the relatively-smaller pre-trained models (\eg{} BART~\cite{chipman2010bart}, BERT~\cite{DBLP:conf/naacl/DevlinCLT19}, and T5~\cite{raffel2020exploring}) used in existing learning-based test generation techniques~\cite{ DBLP:conf/naacl/DevlinCLT19, raffel2020exploring, yang2021generalized, radford2018improving, ramachandran2016unsupervised, dai2015semi}, \chatgpt{} incorporates instruction tuning and RLHF~\cite{bai2022training} (Reinforcement Learning from Human Feedback) on a significantly-larger model, which exhibits better generalization and higher alignment with human intention in various domains. In particular, \chatgpt{} has demonstrated outstanding capability of solving various tasks in code generation and comprehension~\cite{DBLP:journals/corr/abs-2312-10448, DBLP:journals/corr/abs-2311-04448, DBLP:journals/corr/abs-2403-16362, DBLP:journals/corr/abs-2308-01861}. 
To enable a comprehensive evaluation, we first construct a dataset of 1,000 Java focal methods, each along with a complete and executable project environment.  
We incorporate \chatgpt{} to generate unit tests for each focal method and analyze the quality of the generated tests to answer the following four research questions.

\begin{itemize}[topsep=3pt, leftmargin = 13pt]
    \item \textbf{RQ1 (Correctness): How is the correctness of the unit tests generated by \chatgpt{}? }
    We first measure the syntactic correctness, compilation correctness, and execution correctness of the generated tests; and then further build a breakdown of the error types in the incorrect tests. 

    \item \textbf{RQ2 (Sufficiency): How is the sufficiency of the unit tests generated by \chatgpt{}?} We investigate the coverage and assertions of the tests generated by \chatgpt{}.

    \item \textbf{RQ3 (Readability): How is the readability of the unit tests generated by \chatgpt{}?} For those correct tests generated by \chatgpt{}, we perform a user study to assess their readability along with the manually-written tests as reference.

    \item \textbf{RQ4 (Usability): How can the tests generated by \chatgpt{} be used by developers?} For those correct tests generated by \chatgpt{}, we perform a user study to investigate whether developers are willing to adopt them.
\end{itemize}

Based on our results, we have the following main findings. 
On the bad side, we find that only a portion (24.8\%) of tests generated by \chatgpt{} can pass the execution and the remaining tests suffer from diverse correctness issues. 
In particular, 57.9\%  of \chatgpt -generated tests encounter compilation errors, such as using undefined symbols, violating type constraints, or accessing private fields; and 17.3\% of the tests are compilable but fail during execution, which mostly result from the incorrect assertions generated by \chatgpt{}. 
On the good side, we find the passing tests generated by \chatgpt{} actually resemble manually-written ones by achieving comparable coverage, readability, and sometimes even developers' preference compared to manually-written ones. 
Overall, our results indicate that \chatgpt -based test generation could be very promising if the correctness issues in its generated tests could be further addressed. 
To this end, we further distill two potential guidelines, \ie{} providing \chatgpt{} with deep knowledge about the code and helping \chatgpt{} better understand the intention of the focal method, so as to reduce the compilation errors and assertion errors in its generated tests, respectively.

\parabf{Technique.} 
Inspired by our findings above, we further propose \ourtool{}, a novel \chatgpt -based unit test generation approach, which leverages \chatgpt{} itself to improve the correctness of its generated tests. \ourtool{} includes an initial test generator and an iterative test refiner. 
The initial test generator decomposes the test generation task into two sub-tasks by (i) first leveraging \chatgpt{} to understand the focal method via the \textit{intention prompt} and (ii) then leveraging \chatgpt{} to generate a test for the focal method along with the generated intention via the \textit{generation prompt}. 
The iterative test refiner then iteratively fixes the compilation errors in the tests generated by the initial test generator, which follows a validate-and-fix paradigm to prompt \chatgpt{} based on the compilation error messages and additional code context.

To evaluate the effectiveness of \ourtool{}, we further apply \ourtool{} on an evaluation dataset of 100 additional focal methods (to avoid using the same dataset that has been extensively analyzed in our study part) and three projects (to analyze the project-level effectiveness), and compare the tests generated by \ourtool{} with the default \chatgpt{}. 
Furthermore, since the idea of \chatgpt{} is general and not limited to the specific model (\eg{} \chatgpt{}), we further investigate the generalization capabilities of \ourtool{} by applying it to two recent open-source LLMs (\ie{} \codellama{} and \codefuse{}). We answer the following research questions in experiments.

\begin{itemize}[topsep=3pt, leftmargin = 13pt]
    \item \textbf{RQ5 (Improvement): How effective is \ourtool{} in generating correct tests compared to \chatgpt{}? How effective is each component in \ourtool{}?} We compare the number of compilation errors and execution failures between the tests generated by \ourtool{} and the default \chatgpt{}. Moreover, we investigate the contribution of each component in \ourtool{}.

    \item \textbf{RQ6 (Generalization): How effective is \ourtool{} in improving the quality of generated tests when applied to other LLMs?} We replace \chatgpt{} with other LLMs and study the generalization capability of \ourtool{} on helping other LLMs generate high-quality tests. 

    \item \revised{\textbf{RQ7 (Project-level Effectiveness): How effective is \ourtool{} when applied to the entire projects?} We evaluate the coverage, readability, and usability of \ourtool{}-generated tests when applied to entire projects.}


\end{itemize}

Our results show that \ourtool{} substantially improves the correctness of the \chatgpt -generated tests with 34.3\% and 18.7\% improvement in terms of the compilable rate and execution passing rate. 
In addition, our results further confirm the contribution of both components in \ourtool{}, \ie{} the initial test generator is capable to generate more tests with correct assertions while the iterative test refiner is capable to fix the compilation errors iteratively. 
Furthermore, our results show that \ourtool{} can be generalized to the two studied open-source LLMs (\ie{} \codellama{} and \codefuse{}) by  helping both of them generate more correct tests. 

In summary, this paper makes the following contributions:
\begin{itemize}[topsep=3pt, leftmargin = 13pt]
    \item \textbf{The first study} that extensively investigates the correctness, sufficiency, readability, and usability of \chatgpt -generated tests via both quantitative analysis and user study; 

    \item \textbf{Findings and practical implications} that point out the limitations and prospects of \chatgpt -based unit test generation;  
    
    \item \textbf{The first technique \ourtool{}} includes a novel initial test generator and iterative test refiner, which leverages \chatgpt{} itself to improve the correctness of its generated tests;
    
    \item \textbf{An extensive evaluation} that demonstrates the effectiveness and generalization capability of \ourtool{}  by  substantially reducing the compilation errors and incorrect assertions in tests generated by \chatgpt{} and two open-source LLMs. 
\end{itemize}

The data and code can be found on our website~\cite{ChatTESTER}.


\section{Background}

\subsection{Large Language Models}
Large language models (LLMs) are large-scale models pre-trained on a massive textual corpus~\cite{raffel2020exploring, miller1976automatic, ouyang2022training, vaswani2017attention}. In order to fully utilize the massive unlabeled training data, LLMs are often pre-trained with self-supervised pre-training objectives~\cite{DBLP:conf/sigsoft/ChakrabortyADDR22, DBLP:conf/naacl/AhmadCRC21, DBLP:journals/corr/abs-2302-03482}, such as Masked Language Modeling~\cite{feng2020codebert}, Masked Span Prediction~\cite{DBLP:conf/emnlp/0034WJH21}, and Causal Language Modeling~\cite{nijkamp2022codegen}. 
Most LLMs are designed on a Transformer~\cite{vaswani2017attention}, which contains an encoder for input representation and a decoder for output generation. Existing LLMs can be grouped into three categories, including encoder-only (\eg{} CodeBERT~\cite{feng2020codebert}), decoder-only (\eg{} CodeGen~\cite{nijkamp2022codegen}), and encoder-decoder models (\eg{} CodeT5~\cite{DBLP:conf/emnlp/0034WJH21}). 

To enhance the generalization ability and the human-intention-alignment of LLMs on unseen downstream tasks,  more recent work leverages instruction tuning and reinforcement learning to further improve the model performance~\cite{chatGPT, ouyang2022training}. 
For example, \chatgpt{}~\cite{chatGPT}, one of the most representative LLMs developed by OpenAI based on the generative pre-trained transformer (GPT) architecture, first tunes the GPT model with instruction tuning and then updates the model with reinforcement learning from human feedback. 
In addition to commercial LLMs, there are an emerging number of open-source instructed LLMs (\eg{} \codellama{}~\cite{CodeLlama} and \codefuse{}~\cite{CodeFuse}) that also show promising performance in various tasks. 

\subsection{Unit Test Generation}
For a method under test (\ie{} often called as the focal method), unit test generation techniques automatically generate its corresponding unit test, which often consists of a \textit{test prefix} and a \textit{test oracle}~\cite{DBLP:journals/tse/BarrHMSY15}. The test prefix is sequential method invocation statements or assignment statements to drive the focal method to a testable state, and the test oracle (\eg{} assertions) checks whether the focal method behaves consistently with the specification.

Traditional techniques use search-based~\cite{fraser2011evosuite}, random-based~\cite{pacheco2007feedback}, or constraint-based strategies~\cite{ernst2007daikon, csallner2008dysy} to generate unit tests automatically. For example, \evosuite{}~\cite{fraser2011evosuite}, one of the most representative search-based techniques, uses evolutionary algorithms to generate test suites for given Java classes with the goal of maximizing coverage. Although achieving reasonable coverage, the tests generated by traditional techniques have low readability and meaningfulness compared to manually-written ones, which cannot be directly adopted by developers in practice~\cite{almasi2017industrial, grano2018empirical, daka2015modeling, grano2019scented, palomba2016automatic, palomba2016diffusion}. 

Recent work~\cite{tufano2020unit, nie2023learning, watson2020learning, dinella2022toga, mastropaolo2021studying} leverages advanced deep learning techniques, especially LLMs, to generate unit tests. Some 
learning-based techniques regard test generation as a neural machine translation problem, which fine-tunes the model to translate the focal method into the corresponding test prefix or assertion. 
More recently, researchers propose different prompt strategies~\cite{DBLP:conf/icse/NashidSM23, lemieux2023codamosa, kang2022large, ase2023Li} to leverage instructed LLMs to generate test inputs and  assertions.

\section{Study Setup}
\subsection{Benchmark}~\label{sec:study:dataset}
Existing benchmarks on unit test generation ~\cite{ tufano2020unit, watson2020learning} only include a limited code context (\eg{} the focal method alone) rather than a complete and executable project, and it is hard to directly compile and execute the generated tests with the existing datasets. Therefore, to comprehensively evaluate the quality of \chatgpt -generated tests,  we construct a new benchmark of not only focal methods but also complete and executable projects.
 We construct our benchmark as follows. 

\parabf{Project Collection.} We use the 4,685 Java projects in the popular benchmark CodeSearchNet~\cite{husain2019codesearchnet} as the initial project list. 
For each project, we clone it from  GitHub (as of March 25, 2023) and collect its relevant information (\eg{} its creating time and its last commit time). To keep high-quality projects, we filter the 4,685 Java projects according to the following criteria: (i) the project is under continuous maintenance (\ie{} the project should have been updated as of January 1, 2023); (ii) the project has at least 100 stars; (iii) the project is built with Maven framework (for the ease of test executions) and it could be successfully compiled in our local environment.
In this way, we obtain 185 Java projects, containing 502/118 code/test files and  244,876/20,934 lines of code/test code.

\parabf{Data Pair Collection.} We then extract data pairs from the 185 Java projects. Each data pair refers to the pair of the focal method information and its corresponding test method. In particular, in addition to the focal method itself, the focal method information also includes the focal class declaration, all the fields, and all the method signatures (\ie{} the class constructor and the instance methods). For each Java project, we extract data pairs in the following steps. (i) Given a Java project, we first find all the test classes in the project. If a class contains at least one method annotated with \texttt{@Test}, we regard this class as a test class and collect all the test methods in this test class. (ii) We then find the corresponding focal method for each test method based on the file path and the class name matching. For example, for a test method \textit{``testFunction()''} located in the path \textit{``src/test/java/FooTest.java''}, we consider the method \textit{``Function()''} located in the path \textit{``src/main/java/Foo.java''} as its focal method. For the cases when there are multiple focal methods with the same name in the same class, we further filter them by the number and types of parameters to find the unique matching one. For fair comparison, we currently keep the test methods comprising one test case.

With such strict mapping criteria, we extract 1,748 data pairs from 185 Java projects. Considering the costs of using \chatgpt{} API and the manual efforts, we further sample 1,000 data pairs as our final benchmark for the empirical study, which includes test methods and focal methods of diverse scales and structures. The detailed statistics of our benchmark can be found at our website~\cite{ChatTESTER}.

\begin{figure}[htb]
    \centering
    \includegraphics[width=0.8\textwidth]{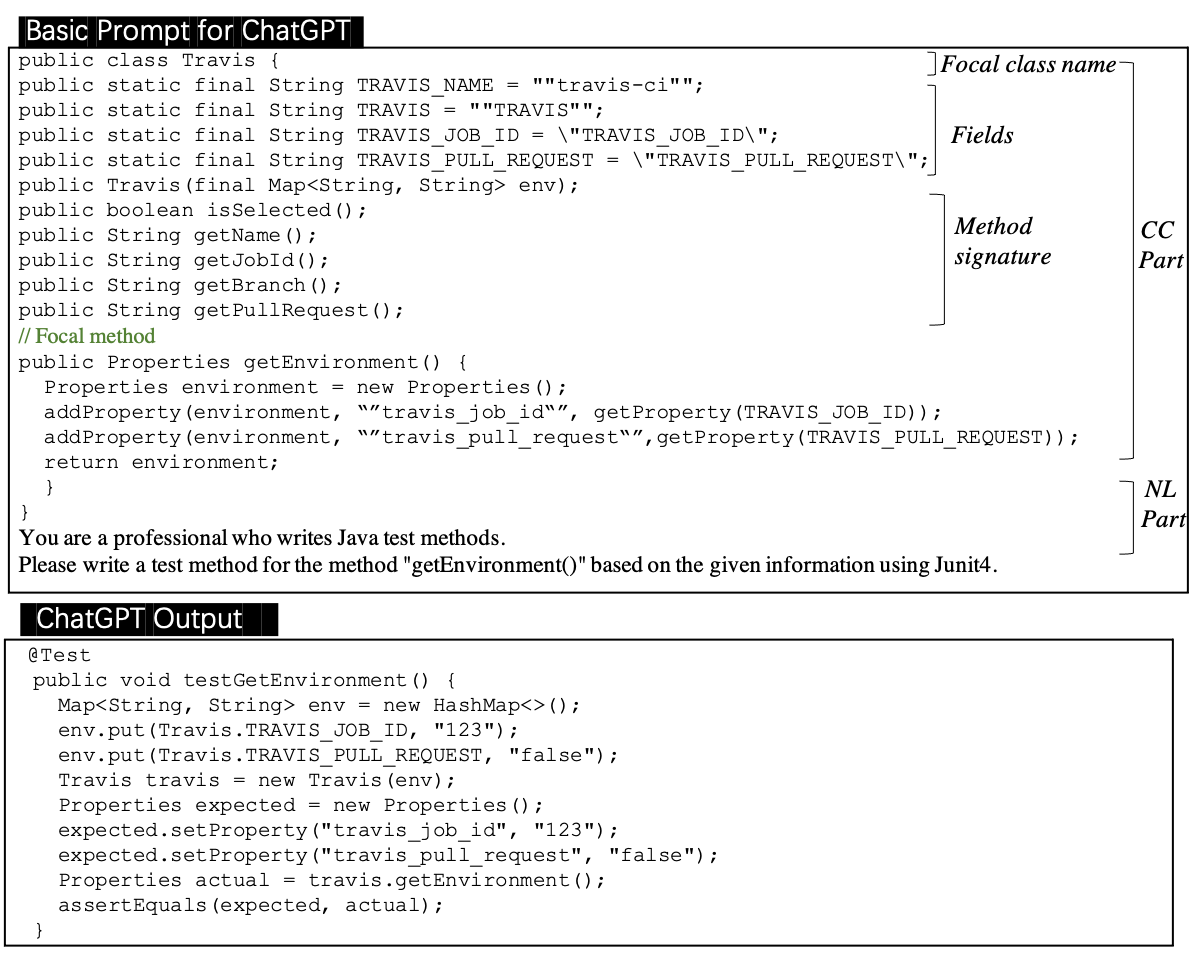}
    \caption{\revised{Basic Prompt for \textit{getEnvironment}~\cite{getEnvironment} Method}}
    \label{figure:basic_prompt}
   
\end{figure}

\subsection{Basic Prompt Design}~\label{sec:study:prompt}
To avoid using too simple prompts that might lead to underestimation of \chatgpt 's capability or using too sophisticated prompts that are uncommon in practice, we design our basic prompt by carefully following the common practice in existing unit test generation work~\cite{tufano2020unit, nie2023learning, watson2020learning} and widely-adopted experience on using \chatgpt{}~\cite{dong2023self,promptStrategies}. In particular, our basic prompt includes two part: (i) the natural language description part (\ie{} NL part) that explains the task to \chatgpt{}, and (ii) the code context part (\ie{} CC part) that contains the focal method and the other relevant code context. We then explain each part in detail. 

\parabf{CC Part.} Following existing learning-based unit test generation work~\cite{tufano2020unit}, we include the following code context into the CC part: (i) the complete focal method, including the signature and body; (ii) the name of the focal class (\ie{} the class that the focal method belongs to); (iii) the field in the focal class; and (iv) the signatures of all methods defined in the focal class. 

\parabf{NL Part.} Based on the widely-acknowledged experience on using \chatgpt{}, we include the following contents in the NL part: (i) a role-playing instruction (\ie{} ``You are a professional who writes Java test methods.'') to inspire \chatgpt 's capability of test generation, which is a common prompt optimization strategy~\cite{dong2023self, promptStrategies}; and (ii) a task-description instruction (\ie{} ``Please write a test method for the \{focal method name\} based on the given information using \{Junit version\}'').

The top half of Figure~\ref{figure:basic_prompt} shows an example of our basic prompt.  After querying with the basic prompt, \chatgpt{} then returns a test as shown in the bottom half of Figure~\ref{figure:basic_prompt}.


\begin{figure}[htb]
    \centering
    \includegraphics[width=0.6\textwidth]{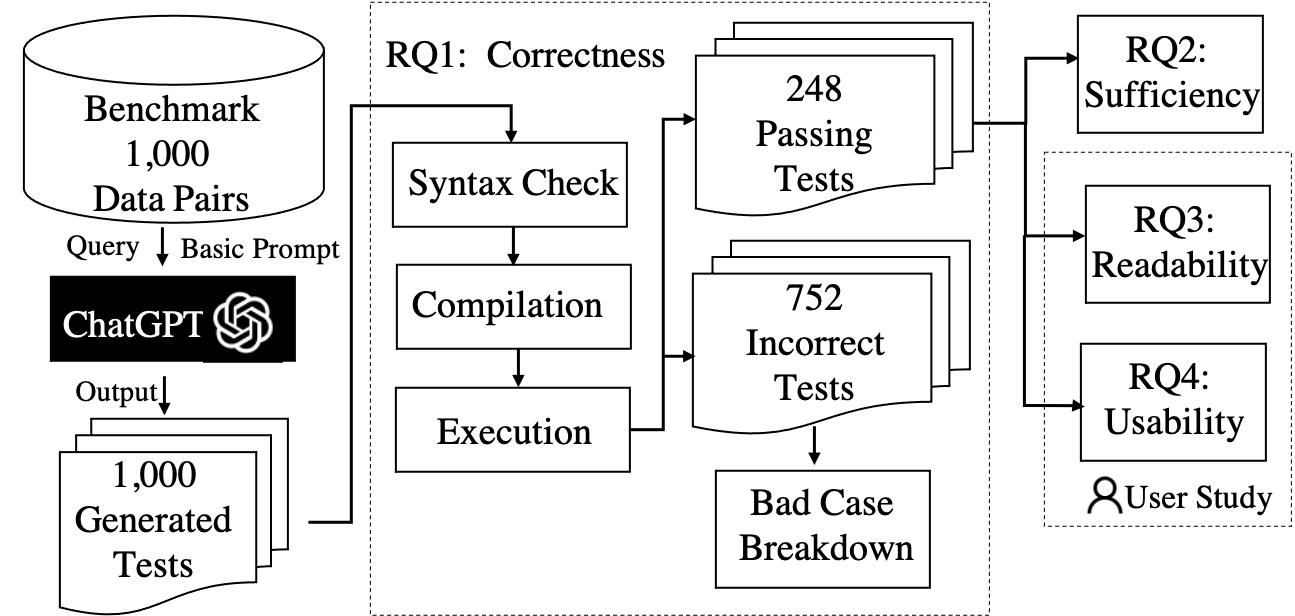}
    \caption{The Workflow of Our Empirical Study} 
    \label{figure:meth}
\end{figure}

\subsection{Baselines}~\label{sec:study:baseline}
We further include two state-of-the-art traditional and learning-based unit test generation techniques as baselines.
We focus on techniques applicable to generating JUnit tests due to our constructed benchmark.
For traditional techniques, we use \evosuite{}~\cite{fraser2011evosuite} as the baseline \revised{with its default setting (\eg{} ``assertion\_strategy'' $=$ MUTATION and ``assertion\_timeout $=$ 60s'')}.
For learning-based techniques, we include \athenatest{}~\cite{tufano2020unit} as the baseline. We do not consider the recent learning-based test completion technique Teco~\cite{nie2023learning} since it mainly targets at statement-level test completion while in this work we focus on directly generating a complete test method. We do not include the two Codex-based test generation techniques (\eg{} CODAMOSA~\cite{lemieux2023codamosa} and LIBRO~\cite{kang2022large}), since they focus on different test generation scenarios as ours (More detailed explanation can be found in Section~\ref{sec:related:test}). 
For \athenatest{}, since it has not released its pre-trained BART-based model or its fine-tuned model, we try our best to reproduce it according to its paper. We reproduce \athenatest{} by fine-tuning the widely-used LLM CodeT5 on the same fine-tuning dataset used in \athenatest{}. We choose CodeT5 since it has been pre-trained on both textual and code corpus, which is also the best pre-training setting shown in the \athenatest{} paper~\cite{tufano2020unit}. 
\revised{In addition, to avoid potential data leakage, we remove the overlap data from the fine-tuning dataset if it is duplicated in our benchmark based on character matching.}

\subsection{Experimental Procedure} \label{section:experimentalProcedure}
Figure~\ref{figure:meth} shows the overview of our experimental procedure. For  each data pair in the benchmark constructed in Section~\ref{sec:study:dataset}, we query \chatgpt{} with our basic prompt designed in Section~\ref{sec:study:prompt} and take the test generated by \chatgpt{} as the output. To automate our experiments, we use the official \chatgpt{} API~\cite{chatGPT} with the default setting. In this work, we focus on the gpt-3.5-turbo model rather than GPT-4, due to the limited rate of GPT-4 API for non-industrial users. We then put the generated test in the same directory of its focal class, and attempt to compile and execute it for further analysis. We then explain the detailed procedure in each RQ, respectively.

\parabf{RQ1: Correctness.} Following existing learning-based test generation work~\cite{tufano2020unit, nie2023learning}, we measure the correctness of the generated tests with three metrics, including (i) syntactic correctness (whether the test could pass the syntax checker), (ii) compilation correctness (whether the test could be successfully compiled), and (iii) execution correctness (whether the test could pass the execution). 
Here we leverage AST parser (such as JavaParser~\cite{javaparser}) as a syntax checker. 
In addition, we further investigate the common error types in those incorrect tests to better understand the limitations of \chatgpt{} in test generation.  
Specifically, we automatically extract the error messages thrown during the compilation and execution, including different compilation and execution error types. 

\parabf{RQ2: Sufficiency.} In this RQ, we include three metrics to assess the sufficiency of the tests generated by \chatgpt{}: (i) the statement coverage of the test on the focal method; (ii) the branch coverage of the test on the focal method; (iii) the number of assertions in the test case.  In particular, we leverage Jacoco~\cite{jacoco} to collect the coverage. 
    
\parabf{RQ3 \& RQ4: User study for readability and usability.} In these two RQs, we conduct a user study to investigate the readability and usability of the tests generated by \chatgpt{}. 
Here, we only focus on 248 passing tests generated by \chatgpt{} since it is less meaningful to recommend tests with compilation errors or execution errors to developers in practice. 
We invite five participants whose Java development experiences vary from 4 years to 5 years. \revised{For each \chatgpt{}-generated test and its corresponding manually-written test, we ask each participant to score their readability and usability via the detailed criteria in Table~\ref{table:criteria}. In particular, the score range of both readability and usability is from 2 to 6, which are the sum of its sub-properties, \ie{} readability consists of the naming intuitiveness and the code layout while the usability consists of the assertion quality and the adoption efforts.}
\revised{Given the large manual efforts required by such an assessment procedure,  we set a reward mechanism (\ie{} 250\$ per participant) and sufficient time (\ie{} two weeks) to mitigate arbitrary responses. In addition, participants are not informed which test is generated by \chatgpt{} or which is written manually.}

\begin{table}[htb]
	\centering
	\small
	\caption{\revised{Scoring Criteria for Readability and Usability}}\label{table:criteria}
	
	\begin{adjustbox}{width=0.95\columnwidth}
	   	
	\begin{tabular}{l|l|l}
        \hline 
        & \multicolumn{2}{c}{\revised{\textbf{Scoring criteria}}} \\ \hline
 
        \multirow{6}{*}{\tabincell{c}{\revised{\textbf{Readability}} \\  \revised{\textbf{(2 \textasciitilde{} 6)}}}}  
        &  \multirow{3}{*}{\tabincell{l}{\revised{\textbf{Naming Intuitiveness (1 \textasciitilde{} 3):}} \\ \revised{the clarity and descriptiveness of} \\ \revised{variable and test method names}}}
         & \revised{3: Most names are intuitive and easy to understand.} \\ \cline{3-3}
        & & \revised{2: Some names are intuitively named and easy to understand.}  \\ \cline{3-3}
        & & \revised{1: Most names are hard to understand without extensive context.} \\ \cline{2-3}

        & \multirow{3}{*}{\tabincell{l}{\revised{\textbf{Code Layout (1 \textasciitilde{} 3):}  the code} \\  \revised{structure, logic, and formatting.}}} 
        & \revised{3: Well-structured code with clear logic.}\\ \cline{3-3}
        & & \revised{2: Overall reasonable structure with minor issues.}  \\ \cline{3-3}
        & & \revised{1: Chaotic logic or massive redundant code} \\ \hline \hline

        \multirow{6}{*}{\tabincell{c}{\revised{\textbf{Usability}} \\  \revised{\textbf{(2 \textasciitilde{} 6)}}}}

        &  \multirow{3}{*}{\tabincell{l}{\revised{\textbf{Assertion Quality (1 \textasciitilde{} 3):}} \\ \revised{validating the focal method with} \\ \revised{appropriate assertions.}}}
        & \revised{3:  Accurate assertions withnin the test context.} \\ \cline{3-3}
        & & \revised{2: Reasonable but weak assertions.} \\ \cline{3-3}
        & & \revised{1: semantically-incorrect assertions.} \\ \cline{2-3}

        &  \multirow{3}{*}{\tabincell{l}{\revised{\textbf{Adoption Efforts (1 \textasciitilde{} 3):}  how} \\ \revised{easily the generated tests can be} \\ \revised{adopted into the practical usage.}}}
        
        & \revised{3: No efforts required before direct adoption.} \\ \cline{3-3}
        & & \revised{2: Simple modifications required before direct adoption.} \\ \cline{3-3}
        & & \revised{1: Extensive efforts required to optimize.} \\ \hline
	\end{tabular}
	\end{adjustbox}
\end{table}

\begin{table}[htb]
	\centering
	\small
	\caption{Correctness of Generated Tests}\label{table:rq1:correctness}
	
	\begin{adjustbox}{width=0.55\columnwidth}

    	\begin{tabular}{l|r|r|r}
		\hline
        \textbf{Metrics (\%)} & \textbf{\chatgpt{}} & \textbf{\athenatest{}} & \textbf{\evosuite{}} \\ \hline  

        {Syntactical correct} & $\approx$ 100.0 & 54.8 &100.0 \\ 
        {Success compilation} & 42.1& 18.8 & \revised{97.6}\\ 
        {Passing execution } & 24.8& 14.4 & \revised{91.4}\\ 
        \hline
	\end{tabular}

	\end{adjustbox}
\end{table}
\vspace{-3mm}

\section{Study Results}
\subsection{RQ1: Correctness}
Table~\ref{table:rq1:correctness} presents the correctness of the  1,000 tests generated by \chatgpt{} and other techniques. Overall, we could observe that a large portion of tests generated by \chatgpt{} suffer from correctness issues, \ie{} 42.1\% of the generated tests are successfully compiled while only 24.8\% of the generated tests are executed successfully without any execution errors. We further manually inspect the failed tests to check whether they actually reveal bugs in the focal method under test, but we find that all of them are caused by the improper test code itself.

As for the learning-based baseline \athenatest{}, \chatgpt{} has a substantial improvement over \athenatest{} in terms of syntactic correctness, compilation correctness, and executable correctness. For example, almost all the tests generated by \chatgpt{} (except the one has an incorrect parenthesis generated) are syntactically correct, but nearly a half of the tests generated by \athenatest{} are syntactically incorrect. The reason might be that the significantly-larger model scale in \chatgpt{} helps better capture syntactical rules in the massive pre-training code corpus. 
As for the traditional search-based baseline \evosuite{}, we could observe a higher compilable rate and passing rate in its generated tests.
In fact, it is as expected since \evosuite{} prunes invalid test code during its search procedure and generates assertions exactly based on the dynamic execution values, while learning-based techniques (\ie{} \chatgpt{} and \athenatest{}) directly generate tests token by token without any post-generation validation or filtering. Therefore, we do not intend to conclude that \evosuite{} is better at generating more correct tests, since the correctness of tests generated by learning-based techniques could be further improved if they also incorporate similar post-generation validation to filter those incorrect tests.

\finding{\chatgpt{} substantially outperforms existing learning-based techniques in terms of syntactic, compilation, and execution correctness. However, only a portion of its generated tests can pass the execution while a large ratio of its generated tests still suffer from compilation errors and execution errors.}

\parabf{Bad Case Breakdown.}
We further analyze the common error types in the failed tests generated by \chatgpt{} (\ie{} those tests failed on the compilation or execution). In particular, we first automatically categorize each test based on the error message; and then we manually summarize and merge similar types into high-level categories. Table~\ref{table:rq1:breakdown1} shows the breakdown for tests with compilation errors, while Table~\ref{table:rq1:breakdown2} shows the breakdown for tests with execution failures.

\begin{table}[htb]

	\centering
	\small
	\caption{Compilation Error Breakdown}\label{table:rq1:breakdown1}
	
	\begin{adjustbox}{width=0.6\columnwidth}
	   	
	\begin{tabular}{c|l|c}
       \hline
         \textbf{Category} & \textbf{Detailed Errors} & \textbf{Frequency} \\   \hline

         \multirow{4}{*}{\tabincell{c}{Symbol  \\ Resolution \\ Error}}  & Cannot find symbol class & 1,934 \\ 
         & Cannot find symbol method & 471 \\ 
         & Cannot find symbol variable & 466 \\ 
         & Cannot find symbol & 169 \\ \hline

        \multirow{3}{*}{Type Error}  & Incompatible types & 73 \\
        & Constructor cannot be applied to given types & 46 \\
        & Methods cannot be applied to given types & 11 \\ \hline 

        \multirow{1}{*}{Access Error} &  Private access  & 75 \\\hline

        \tabincell{c}{Abstract Class \\ Initiation Error} & Abstract class cannot be instantiated & 33 \\ \hline 

        \tabincell{c}{Unsupported \\Operator} & Diamond operator is not supported & 15 \\ \hline

	\end{tabular}
	\end{adjustbox}
\end{table}

\begin{table}[htb]
	\centering
	\small
	\caption{Execution Error Breakdown}\label{table:rq1:breakdown2}
	
	\begin{adjustbox}{width=0.6\columnwidth}
	   	
	\begin{tabular}{c|l|c}
        \hline
        \textbf{Category} &  \textbf{Detailed Errors} & \textbf{Frequency} \\ \hline
      \tabincell{c}{Assertion \\ Error} &  \tabincell{l}{   
      java.lang.AssertionError/ \\ org.opentest4j.AssertionFailedError/ \\org.junit.ComparisonFailure/  \\ org.junit.internal.ArrayComparisonFailure} & 148  \\ \hline

      \multirow{4}{*}{\tabincell{c}{Runtime \\  Error}} & java.lang.IllegalArgumentException &  6 \\ \cline{2-3} 
        &java.lang.RuntimeException	& 5 \\ \cline{2-3}
        & java.lang.NullPointerException &	3 \\  \cline{2-3}
        & others & 11 \\ \hline
     
	\end{tabular}
	\end{adjustbox}
\end{table}

\textit{Failed Compilation.} In Table~\ref{table:rq1:breakdown1}, the column ``Frequency'' shows the number of each compilation errors in 579 uncompilable tests. Note that there could be multiple compilation errors in one test and thus the sum is larger than 579. 
Due to space limits, we only present the frequent compilation errors observed more than 10 times. As shown in the table, the generated tests have diverse compilation errors. 
First, the most frequent compilation errors are caused by un-resolvable symbols, \eg{} the generated tests include some undefined classes, methods, or variables. 
Second, another large category of compilation errors are related to type errors, \eg{} the parameter type in the method invocation is inconsistent with the one defined in the method declaration. 
Third, \chatgpt{} also frequently generates test code that invalidly accesses private variables or methods (\ie{} access \revised{errors).} 
In addition, some generated tests encounter compilation errors by invalidly instating abstract class or using unsupported operators. 

\begin{figure}[htb]
    \centering
    \includegraphics[width=0.7\textwidth]{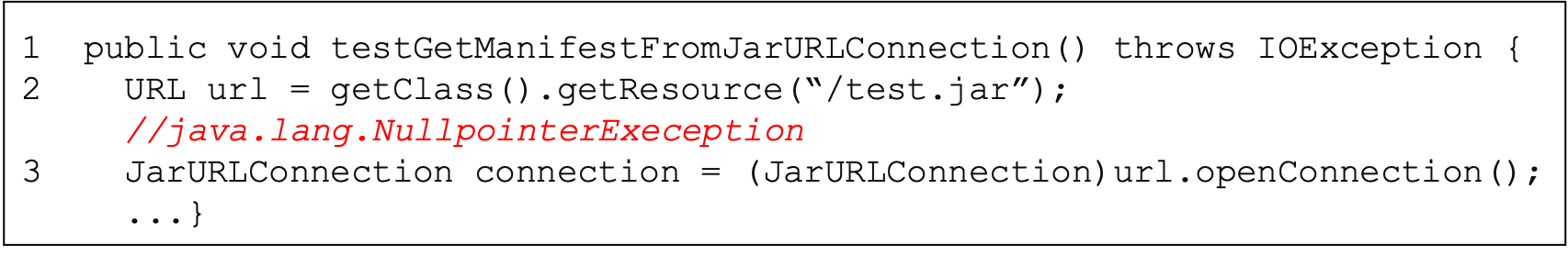}
    \caption{NullPointerException Example}
    \label{fig:exp}
    
\end{figure}

\textit{Failed Execution.} 
In Table~\ref{table:rq1:breakdown2}, we group all the infrequent errors (\ie{} less than three times) into the ``others'' category due to space limits. 
As shown in the table, the majority of failed executions (85.5\%) are caused by assertion errors, \ie{} the assertions generated by \chatgpt{} consider the behavior of the program under test violates the specification. As mentioned above, we manually inspect these assertion errors to identify whether they are caused by the bugs in the focal method or the incorrect assertions themselves, and we find all of them as a result of incorrect assertions generated by \chatgpt{}. It implies that \chatgpt{} might fail to precisely understand the focal method and the quality of the assertions generated by \chatgpt{} should be largely improved. In addition, we observe that the remaining execution errors are related to different runtime exceptions. For example, Figure~\ref{fig:exp} presents an example of the failed execution in the test generated by \chatgpt{}. The test throws \revised{\textit{NullpointerException}} when executing line 3. The error occurs because the created object \textit{``url''} assesses an external resource \textit{``/test.jar''} which does not exist (in line 2). It actually shows an inherent limitation in \chatgpt{}, \ie{} the unawareness of external resources during test generation. 

\finding{The \chatgpt -generated tests encounter diverse compilation errors, such as symbol resolution errors, type errors, and access errors; the majority of failed executions are caused by the incorrectly-generated assertions.}

\subsection{RQ2: Sufficiency}
Table~\ref{table:rq1:sufficiency} presents the statement and branch coverage of generated tests that could pass the execution. As the un-executable tests exhibit zero coverage which can bias the overall results, the coverage is uniformly calculated on the focal methods for which both \chatgpt{} and \evosuite{} can generate executable tests.
We further include the coverage of the \revised{manually-written} tests (\ie{} the original test for the focal method in the project) for reference. 
Since \evosuite{} might generate more than one tests for a focal method, we present the first, worse, average, and union coverage of its generated tests. 
As shown in the table, we could observe that the tests generated \chatgpt{} achieve the highest coverage compared to existing learning-based and search-based techniques and it also achieve comparable coverage as manually-written tests.

\begin{table}[htb]

	\centering
	\small
	\caption{Coverage of Generated Tests}\label{table:rq1:sufficiency}
 
\begin{tabular}{c|c|c|ccccc|c}

\hline
                                &                           &                              & \multicolumn{5}{c|}{Evosuite}                                                                                                                                                                 &                          \\ \cline{4-8}
\multirow{-2}{*}{Coverage (\%)} & \multirow{-2}{*}{ChatGPT} & \multirow{-2}{*}{AthenaTest} & \multicolumn{1}{c|}{First} & \multicolumn{1}{c|}{Worst} & \multicolumn{1}{c|}{Best} & \multicolumn{1}{c|}{Average} &  Union & \multirow{-2}{*}{Manual} \\ \hline
Statement                       & 82.3                      & 65.5                         & \multicolumn{1}{c|}{68.0}  & \multicolumn{1}{c|}{39.7}  & \multicolumn{1}{c|}{76.2} & \multicolumn{1}{c|}{56.9}  & 81.1   & 84.2                     \\ \hline
Branch                          & 65.6                      & 56.2                         & \multicolumn{1}{c|}{61.2}  & \multicolumn{1}{c|}{31.9}  & \multicolumn{1}{c|}{62.1} & \multicolumn{1}{c|}{45.8}   & 77.3 & 68.9  \\ \hline
\end{tabular}
\end{table}

\begin{figure}[htb]
    \centering
    \includegraphics[width=0.5\textwidth]{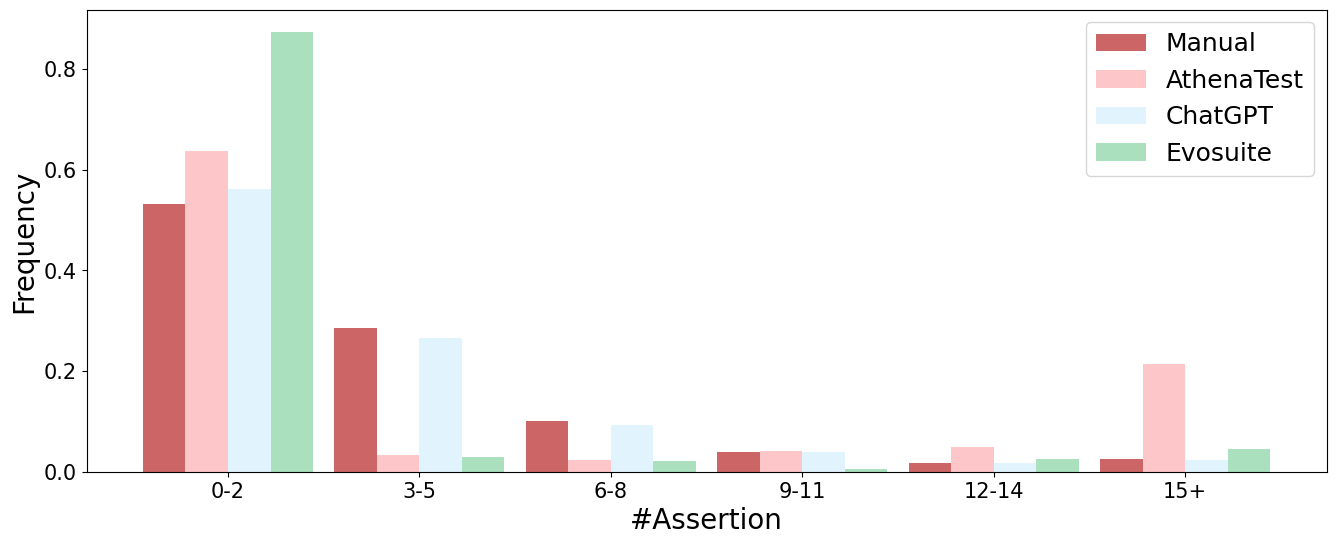}
    \caption{Number of Assertions in Generated Tests}
    \label{fig:assertion}
\end{figure}

Figure~\ref{fig:assertion} presents a distribution plot for the number of assertions in each test generated by different techniques. Interestingly, we observe that \chatgpt -generated tests exhibit most similar distribution as manually-written tests in the number of assertions per test. In particular, \evosuite{} tends to generate tests with less assertions while the learning-based technique \athenatest{} would generate some tests with abnormally-larger number of assertions (\ie{} more than 15 assertions per test) than manually-written ones. The potential reason might be that RLHF helps \chatgpt{} generate more human-like test code. 

\finding{The \chatgpt -generated tests resemble manually-written ones in terms of test sufficiency. \chatgpt{} achieves comparable coverage as manual tests, and also the highest coverage compared to existing techniques;  \chatgpt{} also generate more human-like tests with similar number of assertions per test as manually-written tests.}

\subsection{RQ3: Readability}
\begin{figure}[htb]
    \centering
    \includegraphics[width=0.5\textwidth]{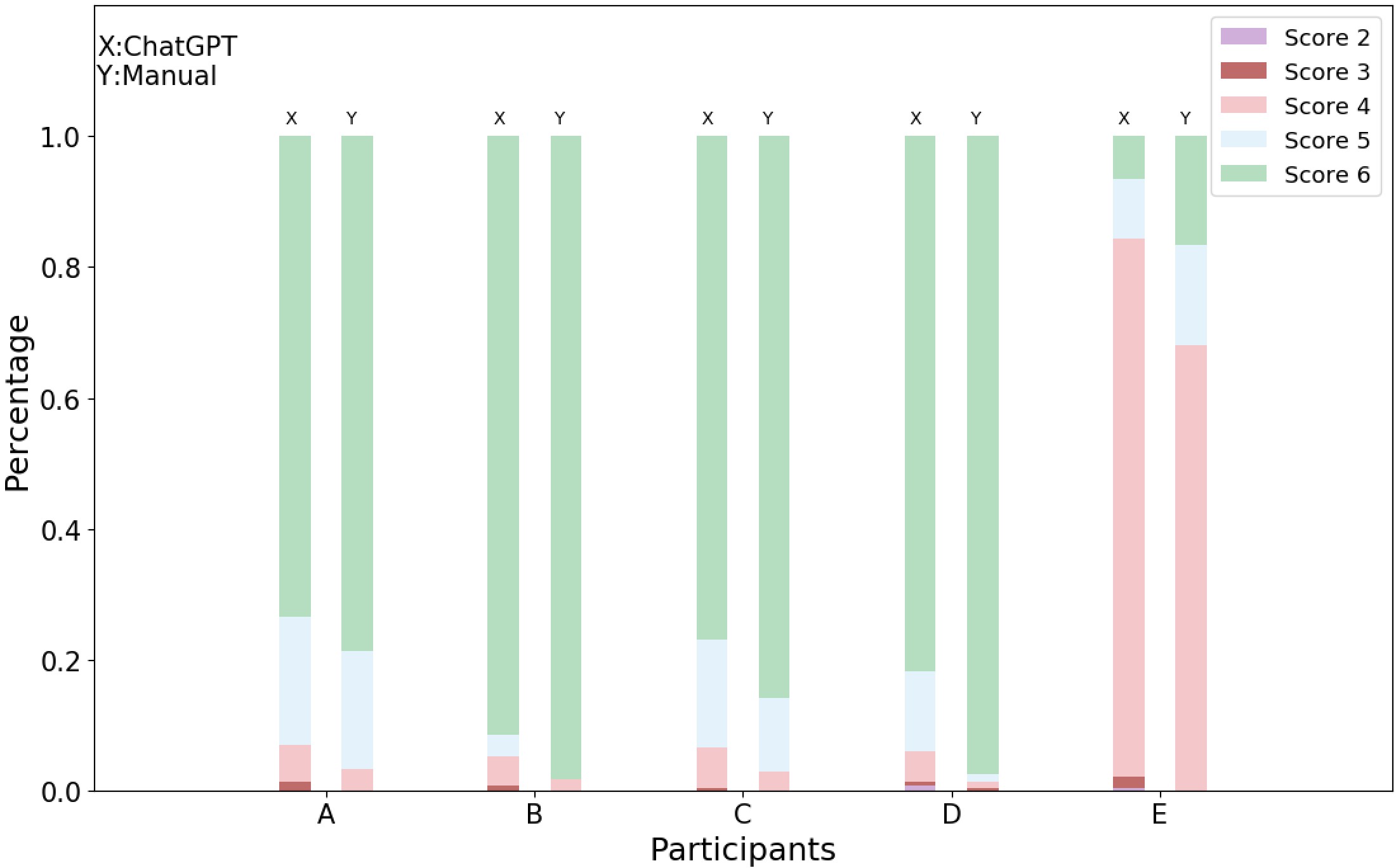}
    \caption{\revised{Response to Readability}}
    \label{fig:user1}
    \vspace{-5mm}
\end{figure}

Figure~\ref{fig:user1} shows the score distribution of readability in a stacked bar chart, where the x-axis represents each participant (\ie{} from A to E) and the y-axis represents the ratio of different scores. Overall, most \chatgpt -generated tests are assessed with decent readability, and they are also considered with comparable and sometimes even better readability compared to manually-written tests. 

\finding{The tests generated by \chatgpt{} have reasonable and comparable readability as manually-written ones. }

\vspace{-4mm}

\subsection{RQ4: Usability}
\begin{figure}[htb]
    \centering
    \includegraphics[width=0.5\textwidth]{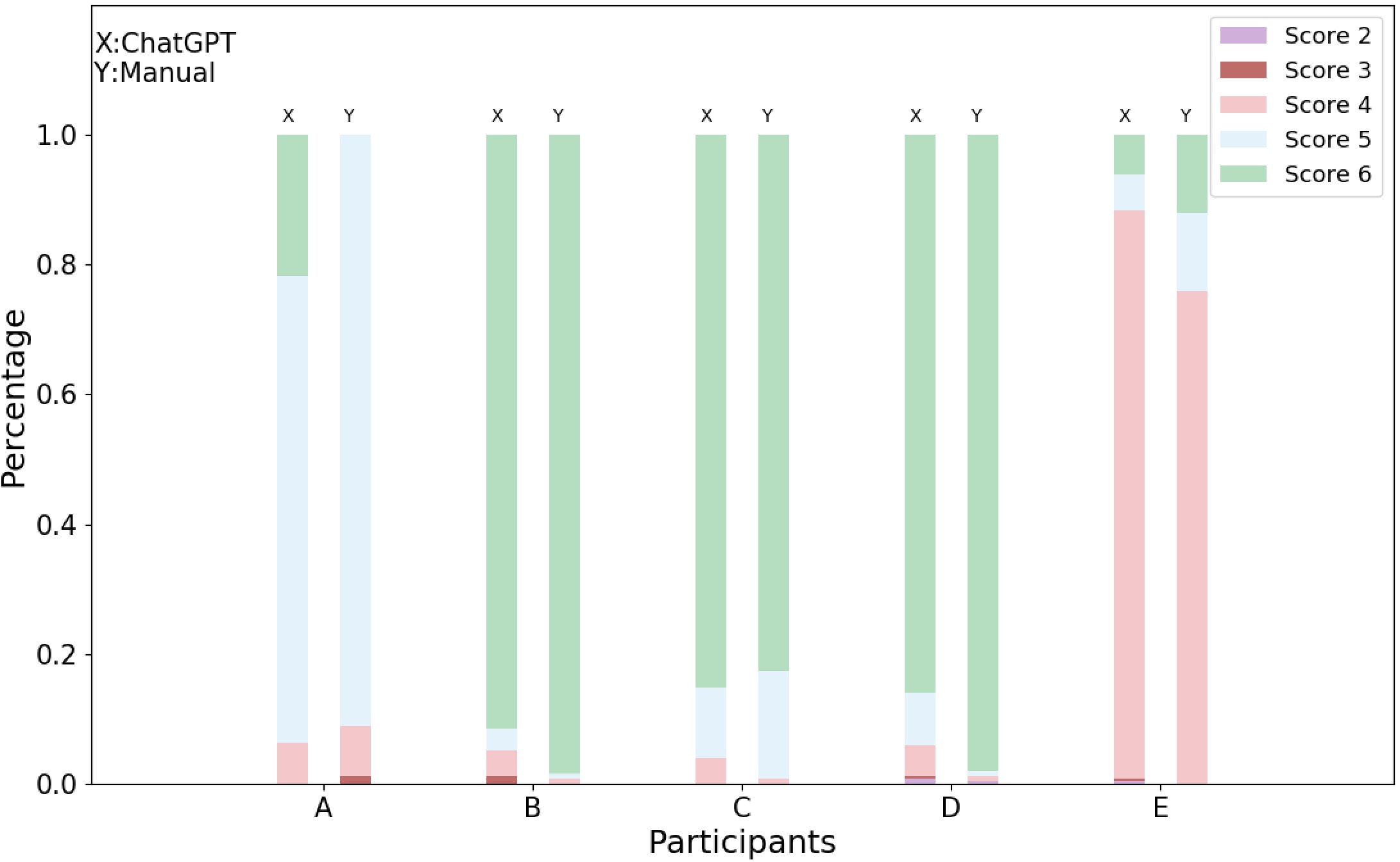}
    \caption{\revised{Response to Usability}}
    \label{fig:user2}
   
\end{figure}

\revised{Figure~\ref{fig:user2} shows the score distribution of usability in a stacked bar chart, where the x-axis represents each participant (\ie{} from A to E) and the y-axis represents the ratio of different scores. Interestingly, we find the \chatgpt{} -generated tests exhibit comparable usability compared to manually-written ones. Based on the scoring details of the two sub-properties in usability (\ie{} assertion quality and adoption efforts, we find that most \chatgpt{}-generated tests are assessed with high-quality assertions for immediate usage, which makes the participants' high willing of direct usage.}

\finding{\chatgpt -generated tests show a large potential in practical usability. In a considerable portion of cases, participants are willing to directly adopt \chatgpt -generated tests.}


\vspace{-4mm}

\subsection{Enlightenment}~\label{sec:rq1:dis}
Based on our results above, we further discuss the implications on the strengths and the limitations of \chatgpt -based unit test generation.

\parabf{Limitations.} As shown in our results, a large portion of \chatgpt -generated tests fail in compilation or execution, which might result from two inherent limitations in  generative language models.

First, most compilation errors might be caused by \chatgpt 's unawareness of the ``deep knowledge'' in the code. Although being pre-trained on a massive code corpus could help \chatgpt{} capture the syntactical rules in the code, the nature of \chatgpt{} is still a probabilistic token-sequence generator and thus it is challenging for \chatgpt{} to be fully aware of the deep rules in the code, \eg{} only the public fields could be accessed outside the class and the abstract classes cannot be instantiated. \textit{Therefore, to help \chatgpt{} overcome this limitation, it is important to remind \chatgpt{} of such deep knowledge during its generating tests.}

Second, most execution errors (\ie{} assertion errors) result from \chatgpt 's lack of understanding about the intention of the focal method. As a result, it is challenging for \chatgpt{} to write proper assertions as specification for the focal method under test. \textit{Therefore, to help \chatgpt{} overcome this limitation, it is essential to help \chatgpt{} to better understand the intention of the focal method.}

\parabf{Strengths.} Although generating a lot of tests failed in compilation or execution, the good thing is that most of the passing tests generated by \chatgpt{} are often of high quality in terms of the sufficiency, the readability, and the usability in practice. These passing tests could mostly be put into direct use to alleviate manual test-writing efforts. \textit{Therefore, leveraging \chatgpt{} to generate unit tests is a promising direction if the correctness issues in its generated tests could be further addressed.}

 	\vspace{1mm}
	\begin{mdframed}[linecolor=gray,roundcorner=12pt,backgroundcolor=gray!15,linewidth=3pt,innerleftmargin=2pt, leftmargin=0cm,rightmargin=0cm,topline=false,bottomline=false,rightline = false]
	\textbf{Enlightenment:} \chatgpt -based unit test generation is promising since it is able to generate a number of high-quality tests with comparable sufficiency, readability, and usability as manually-written tests. However, further efforts are required to address the correctness issues in the \chatgpt -generated tests. The two directions to this end are (i) to provide \chatgpt{} with deep knowledge about the code and (ii) to help \chatgpt{} better understand the intention of the focal method, so as to reduce its compilation errors and assertion errors, respectively.
	\end{mdframed}
	\vspace{1mm}


\section{Approach of \ourtool{}} \label{sec:Approach}

\begin{figure}[htb]
    \centering
    \includegraphics[width=0.6\textwidth]{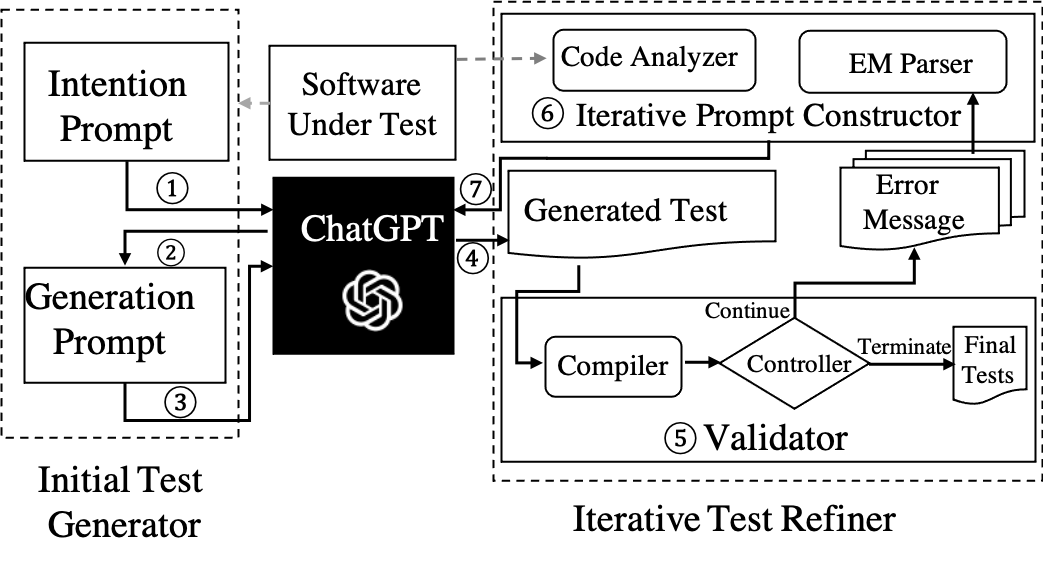}
    \caption{The Workflow of \ourtool{}}
    \label{figure:tech}
    
\end{figure}

\parabf{Overview.} Inspired by our findings and enlightenment above, we then propose \ourtool{}, a novel \chatgpt -based unit test generation approach, which improves the correctness of \chatgpt -generated tests by \chatgpt{} itself.  In particular, \ourtool{} contains two components, \ie{} an initial test generator and an iterative test refiner. Figure~\ref{figure:tech} shows the workflow of \ourtool{}.

Instead of directly asking \chatgpt{} to generate a test for the given focal method, the initial test generator decomposes the test generation task into two sub-tasks: (i) first understanding the intention of the focal method, and (ii) then generating a unit test for the focal method based on the intention. 
Compared to the basic prompt, the initial test generator aims to generate tests with higher-quality assertions based on the help of the intermediate step of intention generation.

The iterative test refiner iteratively fixes the compilation errors in the tests generated by the initial test generation. As mentioned in our enlightenment, the key to eliminating most uncompilable tests is to provide ``deep knowledge'' to \chatgpt{} during test generation. 
However, given the large number of such potential rules, it is infeasible to include all of them in the prompt in advance to the test generation. 
Therefore, we follow a validate-and-fix paradigm to iteratively refine the uncompilable test by prompting \chatgpt{} with compilation error messages and additional relevant code context. In other words, the iterative test refiner actually leverages the error messages from the compiler as the violation instances of the ``deep knowledge'',  so as to fix compilation errors in the generated tests.

\begin{figure}[htb]
    \centering
    \includegraphics[width=0.7\textwidth]{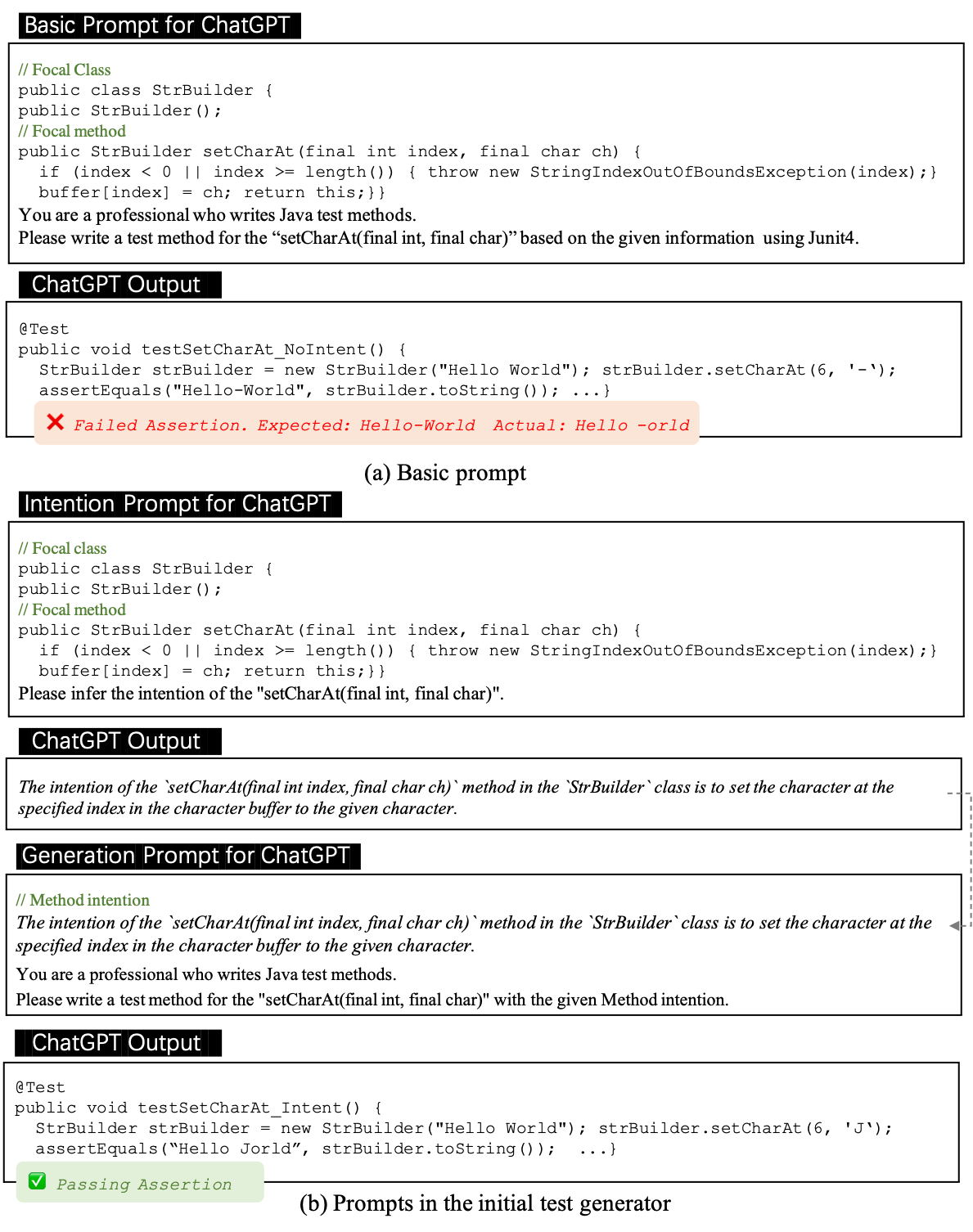}
    \caption{Basic Prompt v.s. Prompts in the Initial Test Generator}
    \label{figure:initial_prompt}
\end{figure}

\subsection{Initial Test Generator}
The initial test generator decomposes test generation into two steps: (i) first leveraging \chatgpt{} to understand the focal method via the \textit{intention prompt}, and (ii) then leveraging \chatgpt{} to generate a test for the focal method along with the generated intention via the \textit{generation prompt}. 

The intention prompt asks \chatgpt{} to return a natural language description of the intended functionality of the focal method under test. In particular, the code context part is similar to the basic prompt in Section~\ref{sec:study:prompt}, including the class declaration, constructor signatures, relevant fields, and the focal method itself;
and the natural language instruction is to ask \chatgpt{} to infer the intention of the focal method. Then, the generation prompt further includes the generated intention and asks \chatgpt{} to generate a unit test for the focal method.

Figure~\ref{figure:initial_prompt} presents an example comparing how the basic prompt and the initial test generator generate a test for the same given focal method \textit{``setCharAt()''}. 
As shown in Figure~\ref{figure:initial_prompt} (a), given the basic prompt without any intention inference, \chatgpt{} generates a test with an incorrect assertion (\ie{} \textit{``assertEquals(Hello-World, strBuilder.toString())''}.
However, in Figure~\ref{figure:initial_prompt} (b), with the intention prompt, \chatgpt{} first correctly generates the intention for the focal method \textit{``setCharAt()''}; and then with the generation prompt, \chatgpt{} generates a test with a correct assertion (\ie{} \textit{``assertEquals(Hello Jorld, strBuilder.toString())''}. The additional intention inference is designed to enhance \chatgpt 's understanding about the focal method, which further leads to more accurate assertion generation.

\begin{figure}[htb]
    \centering
    \includegraphics[width=0.7\textwidth]{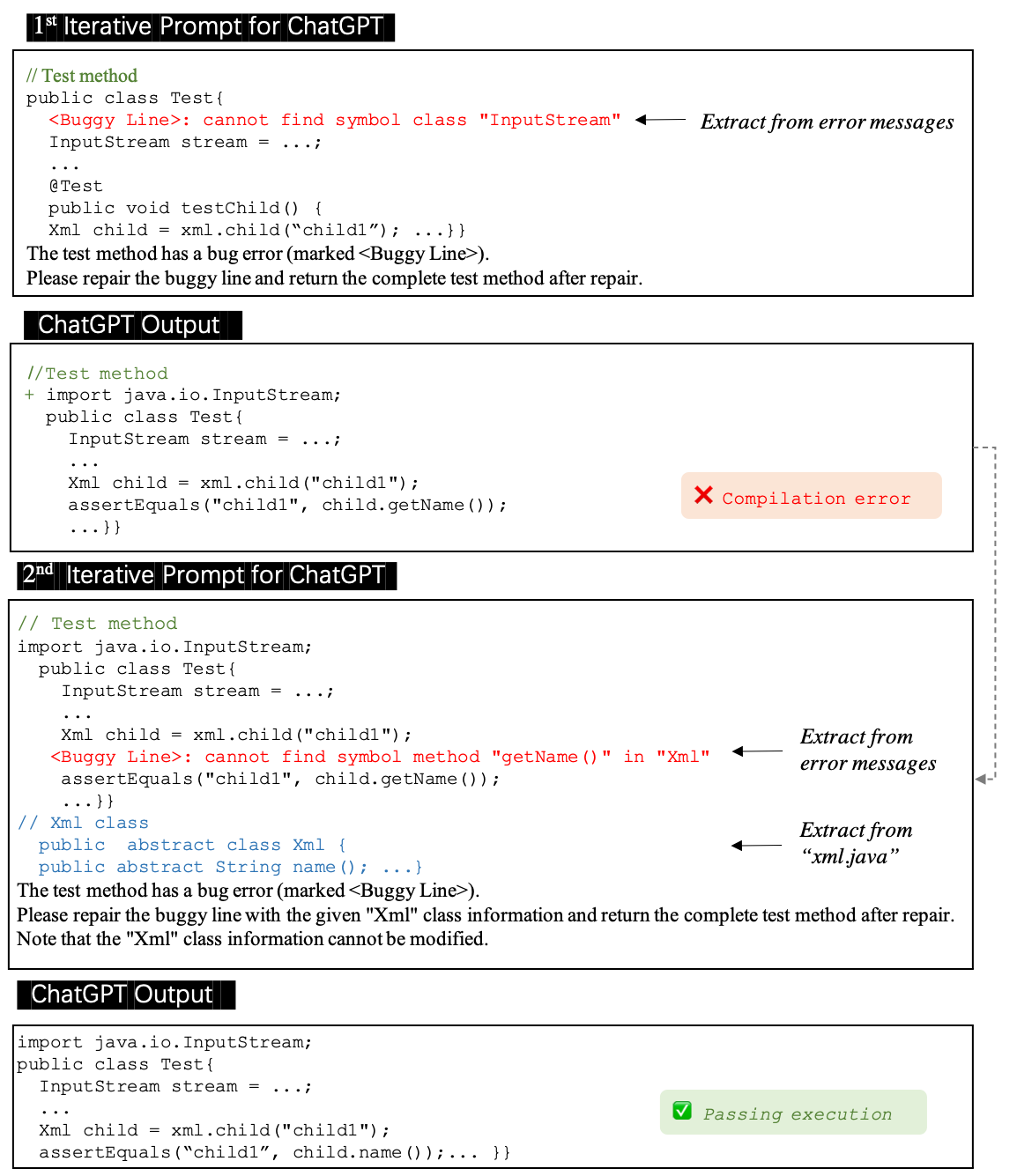}
    \caption{Prompt in the Iterative Test Refiner}
    \label{figure:repair_prompt}
\end{figure}

\vspace{-1mm}
\subsection{Iterative Test Refiner}
The iterative test refiner iteratively fixes the compilation errors in the tests generated by the initial test generation. Each iteration successively leverages two steps: (i) first validating the generated test by compiling it in a validation environment; (ii) second constructing a prompt based on the error message during compilation and the extra code context related to the compilation error. The new prompt is then queried into \chatgpt{} to get a refined test. Such a procedure repeats until the generated test can be successfully compiled or the maximum number of iterations is reached. Currently, we only focus on fixing compilation errors instead of execution errors, since in practice it is challenging to identify whether a test execution failure is caused by the incorrect test code or by the incorrect focal method.  
We then explain each step with the illustration example in Figure~\ref{figure:repair_prompt}.

\parabf{Validator.} For ease of compiling the generated test, we directly create a test file in the same directory of the focal class. In particular, the generated test method is encapsulated in a test class with relevant import statements. Then, the test file is compiled with the Java compiler. A controller then decides the next step based on the compilation status:

\begin{itemize}[topsep=3pt, leftmargin = 13pt]
    \item \textit{Successful compilation:} if there is no compilation error, the controller would terminate the iterative refinement procedure and return the final test;

    \item \textit{Valid refinement:} if the number of compilation errors is less than that in the last iteration, the current refinement is considered as a valid refinement. The controller then proceeds to the iterative prompt constructor so as to continue the refinement;

    \item \textit{Invalid refinement:} if the number of compilation errors is larger than or the same as that in the last iteration, the current refinement is considered an invalid refinement. The controller would terminate the refinement if the accumulated number of invalid refinements is larger than the maximum (\eg{} 3 in our experiments); or proceeds to the iterative prompt constructor.     
\end{itemize}

\parabf{Iterative Prompt Constructor.}
The iterative prompt constructor is built on top of (i) an EM parser that analyzes the error message about the compilation error, and (ii) a code analyzer that extracts the additional code context related to the compilation error. 

In particular, the EM parser collects three types of information by parsing the error message:

\begin{itemize}[topsep=3pt, leftmargin = 13pt]

    \item \textit{Error type}: the high-level description about the error, which is often the first sentence in the error message. For example, \textit{``cannot find simple class ...''} and \textit{``cannot find symbol method ...''} are extracted error types in the illustration example.
    
    \item \textit{Buggy location}: the line number of the test code triggering compilation errors. With such location information, the prompt constructor is able to insert the relevant information around the buggy line, \ie{} starting with the tag \textit{``$\langle$Buggy line$\rangle$''} as shown in the example. 

    \item \textit{Buggy element}: the objects or variables in the buggy location. For example, for the second iteration in Figure~\ref{figure:repair_prompt}, we analyze the buggy line with the error message, and find that they are associated with the class \textit{``Xml''} (which is the buggy element in this example).

\end{itemize}

With the buggy elements, the code analyzer is then able to extract additional code context from other Java files rather than the focal class. In particular, the code analyzer first parses the whole project to find the class file that the buggy element belongs to, and then extracts the class declaration and public method signature from the class file.  This extracted class information would further be added to the prompt as additional information, \eg{}  \textit{``//Xml class ...''} highlighted in blue. In fact, such additional information from other classes could be very important for generating high-quality tests, since it is very common that the test code involves not only the focal class but also other classes. However, given the limited input length for \chatgpt{}, it is infeasible to directly include the whole program in the prompt (which would also lead to bad performance since the focus of \chatgpt{} might be confused). Therefore, in \ourtool{}, we propose to append necessary additional code contexts in such an iterative way.

\section{Evaluation of \ourtool{}}
We systematically evaluate \ourtool{} for its effectiveness (RQ5), its generalization capability on different LLMs (RQ6), and \revised{its project-level effectiveness (RQ7)}. 

\subsection{Evaluation Setup}
\parabf{RQ5 Setup.} As for the evaluation dataset in RQ5, we re-sample another 100 data pairs from the remaining 748 data pairs in Section~\ref{sec:study:dataset} (recall that we collect 1,748 data pairs in total and 1,000 of them are included in the benchmark for the empirical study). The reason that we construct such an additional evaluation dataset for evaluation is to avoid using the same benchmark that has been extensively analyzed in our previous study. Since our approach is inspired by the findings from our study, evaluating it on a separate dataset could eliminate the potential overfitting issues. 

As for the studied techniques in RQ5, to evaluate the overall effectiveness of \ourtool{} and the individual contribution of each component (\ie{} the initial test generator and the iterative test refiner) in \ourtool{}, we study four techniques (1) {\chatgpt{}}: the default \chatgpt{} with the basic prompt, which is the one used in our empirical study; (2) {\ourtoolm{}}: a variant of \ourtool{} without the iterative test refiner, which actually enhances the default \chatgpt{} with the initial test generator of \ourtool{}; \revised{(3) {\ourtooIte{}}: a variant of \ourtool{} without the initial test generator, which actually enhances the default \chatgpt{} with the iterative test refiner of \ourtool{};} (4) \ourtool{}: the complete \ourtool{} with both the initial test generator and the iterative test refiner. 
To mitigate the randomness in \chatgpt{}, we repeat all experiments three times and present the average results.

\parabf{RQ6 Setup.} To evaluate the generalization capability of \ourtool{} with different LLMs, we replace \chatgpt{} in \ourtool{} with two different open-source LLMs, \ie{} \codellama{}-34B~\cite{CodeLlama} and \codefuse{}-34B~\cite{CodeFuse}. For both studied LLMs, we compare their effectiveness of test generation with (i) the basic prompt in a default way and (ii) with the \ourtool{} framework.

\parabf{\revised{RQ7 Setup.}} \revised{To evaluate the project-level effectiveness of \ourtool{},  we select three representative projects from the 185 Java projects in Section~\ref{sec:study:dataset}. Table~\ref{table:projectInfo} presents the characteristics of the selected projects, which belong to different domains with 3k to 6k lines of code.} 
\revised{We apply \ourtool{} and all baselines (\ie{} \athenatest{}, \chatgpt{}, and \evosuite{}) to generate tests for each method in the project. To evaluate the test sufficiency in such a project-level setting, we further compare the statement coverage and branch coverage of the generated tests on each entire project. 
For fair comparison, we randomly select one test when \evosuite{} generates multiple tests for one focal method. 
In addition, we further perform a user study to manually compare the readability and usability of the tests generated by \ourtool{} and baselines on a statistically-sampled subset (\ie{} 219 tests for each technique, which are sampled within the 0.05 error margin at a 95\% confidence level~\cite{singh2013elements, wackerly2014mathematical}). We follow the similar user study methodology in Section~\ref{section:experimentalProcedure}, \ie{} the five participant are asked to score the readability and usability of the generated tests based on the scoring details in Table~\ref{table:criteria}.}

\begin{table}[htb]

	\centering
	\small
	\caption{\revised{Overview of Projects in RQ7}}\label{table:projectInfo}
 \begin{adjustbox}{width=0.6\columnwidth}

\begin{tabular}{c|
c|
c}
\hline
Project Name                                        & \# Line Of Code & Domain                                 \\ \hline
zappos-json~\cite{zappos-json}                                 & 3,552              & Serialization                          \\ \hline

tabula-java~\cite{tabula-java}      & 5,586               & { File processing} \\ \hline
jInstagram~\cite{jInstagram} & 6,303         & Instagram API wrapper                            \\ \hline
\end{tabular}
\end{adjustbox}
\end{table}

\begin{table}[htb]
	\centering
	\small
	\caption{Effectiveness of \ourtool{}}\label{table:rq4}
	
	\begin{adjustbox}{width=0.7\columnwidth}
	   	
	\begin{tabular}{l|r|r|r|r}
		\hline
        \textbf{Metrics (\%)} & \textbf{\chatgpt{}}  &  \revised{\textbf{\ourtoolm{}}} & \revised{\textbf{\ourtooIte{}}} & \textbf{\ourtool{}} \\ \hline  
    {Syntactical correct} & 100.0 & 100.0 & 
    \revised{100.0} & 100.0 \\ 
        {Success compilation} & 39.0 & 50.7 & \revised{60.6} & 73.3\\ 
        {Passing execution} & 22.3 & 29.7 & \revised{34.0} & 41.0\\ 
        \hline
	\end{tabular}
	\end{adjustbox}

\end{table}

\subsection{Evaluation Results}

\subsubsection{RQ5: Effectiveness Evaluation}
Table~\ref{table:rq4} presents the correctness of the tests generated by \chatgpt{} and our approaches. 
Overall, we could observe a substantial improvement in both the compilation rate and passing rate of the tests generated by \ourtool{} compared to the default \chatgpt{}. 
For example, additional 34.3\% tests (= 73.3\% - 39.0\%) can be successfully compiled and additional 18.7\% tests (= 41.0\% - 22.3\%) can pass the execution. 
In summary, the proposed approach \ourtool{} effectively improves the correctness of the tests generated by \chatgpt{}. 

In addition, we could observe that the variant \ourtoolm{} outperforms the default \chatgpt{} by achieving 11.7\% and 7.4\% improvements in the compilation rate and passing rate.
In particular, we find that among the \chatgpt -generated tests with incorrect assertions, 12.5\% of them are fixed into correct assertions in \ourtoolm{}, indicating the effectiveness of the initial test generator. Figure~\ref{figure:initial_prompt} is an example of how the intention prompt improves the correctness of assertions in our dataset. \revised{In addition, when comparing \ourtool{} against the variant \ourtooIte{}, we can find that removing the initial test generator would decrease 12.7\% compilation rate and 7.0\% passing rate, further confirming the contribution of the initial test generator.}
Moreover, we could observe a further improvement from \ourtoolm{} to \ourtool{}, \ie{} additional 22.6\% tests and 11.3\% tests are fixed by the iterative test refiner into compilable tests and passing tests. Figure~\ref{figure:repair_prompt} is an example of how the iterative test refiner fixes the compilation errors in two iterations. 
In summary, both components (\ie{} the initial test generator and the iterative test refiner) positively contribute to the effectiveness of \ourtool{}.

\begin{table}[htb]
    \centering
	\small
	\caption{Distribution of Iteration Numbers in \ourtool{}}\label{table:iteration}
 
    \begin{adjustbox}{width=0.6\columnwidth}
        \begin{tabular}{c|c|c|c|c|c|c|c|c|c|c|c}
        \hline
        \textbf{\# Iteration}  & \textbf{0}  & \textbf{1}  & \textbf{2} & \textbf{3} & \textbf{4} & \textbf{5}  & \textbf{6}  & \textbf{7}  & \textbf{8} & \textbf{9} & \textbf{10} \\ \hline
        \textbf{\# Tests}         & 37 & 12 & 9 & 0 & 1 & 12 & 18 & 10 & 0 & 0 & 1  \\ \hline
        \textbf{\# Success Compilation} & 31 & 10 & 6 & 0 & 1 & 0  & 15 & 6  & 0 & 0 & 0  \\ \hline
        \textbf{\# Passing Execution}   & 31 & 9  & 2 & 0 & 0 & 0  & 0  & 0  & 0 & 0 & 0  \\ \hline
        \end{tabular}
    \end{adjustbox}
\end{table}
\parabf{Costs and number of iterations.}
Table~\ref{table:iteration} presents the distribution of iterative test refinement times of \ourtool{}. The row ``\# Iteration'' shows the number of iterations; the row ``\# Tests '' shows the number of tests that are still being refined in this iteration; the row ``\# Success Compilation'' and the row ``\# Passing Execution'' show the number of tests that are successfully compiled and executed in the current iteration, respectively. Based on the table, we find that the iterative refinement continuously improves the quality of the generated tests.
\revised{The average time cost of \ourtool{} is around 99.0 seconds,} where the initial test generator takes 15.0 seconds while each iteration in the iterative refiner takes 30.0 seconds on average. In fact, the majority of the costs come from the latency of querying the \chatgpt{} API \revised{(\ie{} around 10s in our environment)} and the test compilation/execution. Therefore, the efficiency of \ourtool{} could be further improved if it is applied on locally-deployed small LLMs or with more efficient compilation optimization (\eg{} incrementally compiling only the generated test instead of compiling the whole project).

\vspace{2mm}
\begin{mdframed}[linecolor=gray,roundcorner=12pt,backgroundcolor=gray!15,linewidth=3pt,innerleftmargin=2pt, leftmargin=0cm,rightmargin=0cm,topline=false,bottomline=false,rightline = false]
\textbf{Effectiveness Evaluation Summary:} \ourtool{} effectively improves the correctness of \chatgpt -generated tests by substantially reducing the compilation errors and incorrect assertions in the generated tests. In particular, both the initial test generator and the iterative test refiner positively contribute to the effectiveness of \ourtool{}.
\end{mdframed}

\begin{table}[htb]
    \centering
	\small
	\caption{Generalization of \ourtool{}}\label{table:rq5}
 
    \begin{adjustbox}{width=0.8\columnwidth}
        \begin{tabular}{c|cc|cc}
        \hline
        \multirow{2}{*}{\textbf{Metric (\%)}}   & \multicolumn{2}{c|}{\textbf{CodeLlama-34B}}                              & \multicolumn{2}{c}{\textbf{CodeFuse-34B}}                             \\ \cline{2-5} 
                                 & \multicolumn{1}{c|}{$CodeLLama$} & $CodeLlama_{\ourtool{}}$ & \multicolumn{1}{c|}{$CodeFuse$} & $CodeFuse_{\ourtool{}}$ \\ \hline
        Syntactical Correct                   & \multicolumn{1}{c|}{93.0}           & 93.0                  & \multicolumn{1}{c|}{72.0}          & 89.0                 \\ \hline
        Success Compilation          & \multicolumn{1}{c|}{22.0}           & 43.0                  & \multicolumn{1}{c|}{1.0}           & 24.0                 \\ \hline
        Passing execution          & \multicolumn{1}{c|}{3.0}            & 21.0                  & \multicolumn{1}{c|}{0.0}           & 11.0                 \\ \hline
        \end{tabular}
    \end{adjustbox}

\end{table}

\subsubsection{RQ6: Generalization Evaluation}
Table~\ref{table:rq5} presents the effectiveness of \ourtool{} on different LLMs. Overall, \ourtool{} can consistently improve the quality (\eg{} syntactical, compilation, or execution correctness) of the tests generated by different open-source LLMs. For the studied LLMs \codellama{}-34B and \codefuse{}-34B, \ourtool{} achieves 21.0\%/ 23.0\% improvement in compilation rate and 11.0\% /18.0\% improvement in passing execution rate. The results indicate that the approach of \ourtool{} can be generalized to different LLMs and is not specific to the commercial LLM \chatgpt{}. 
\revised{The average time costs of \codellama{}, $CodeLlama_{\ourtool{}}$, \codefuse{} and $CodeFuse_{\ourtool{}}$ are 5s, 105s, 5s and 90s.}

\vspace{2mm}
\begin{mdframed}[linecolor=gray,roundcorner=12pt,backgroundcolor=gray!15,linewidth=3pt,innerleftmargin=2pt, leftmargin=0cm,rightmargin=0cm,topline=false,bottomline=false,rightline = false]
\textbf{Generalization Evaluation Summary:} \ourtool{} can be generalized to different LLMs and consistently improve their effectiveness of test generation.
\end{mdframed}
\vspace{2mm}

\begin{table}[]

	\centering
	\small
	\caption{\revised{Statement and Branch Coverage in Project-level Evaluation }}\label{table:projectcoverage}

\begin{tabular}{
c |
c 
c |
c 
c |
c 
c }
\hline
                                & \multicolumn{2}{c|}{zappos-json}   & \multicolumn{2}{c|}{{\color[HTML]{262626} tabula-java}} & \multicolumn{2}{c}{jInstagram}  \\ \cline{2-7} 
\multirow{-2}{*}{Coverage (\%)} & \multicolumn{1}{c|}{SC}   & BC    & \multicolumn{1}{c|}{SC}               & BC              & \multicolumn{1}{c|}{SC}   & BC   \\ \hline
AthenaTest                                              & \multicolumn{1}{c|}{7.2}  & 1.3  & \multicolumn{1}{c|}{3.2}              & 0.4             & \multicolumn{1}{c|}{18.9} & 1.6  \\ \hline
EvoSuite                                         & \multicolumn{1}{c|}{30.5} & 16.1  & \multicolumn{1}{c|}{15.2}              & 9.9            & \multicolumn{1}{c|}{45.3} & 13,3  \\ \hline
ChatGPT                                                 & \multicolumn{1}{c|}{14.4} & 5.2  & \multicolumn{1}{c|}{9.4}              & 5.0             & \multicolumn{1}{c|}{38.6} & 12.7 \\ \hline
ChatTester                                              & \multicolumn{1}{c|}{35.1} & 21.5 & \multicolumn{1}{c|}{15.6}             & 9.2             & \multicolumn{1}{c|}{41.8} & 13.4 \\ \hline
\end{tabular}
\end{table}

\subsubsection{\revised{RQ7: Project-level Evaluation}} \revised{We present the results of coverage comparison and the user study results of readability and usability as follows.}

\parabf{\revised{Project-level Coverage.}}
\revised{Table~\ref{table:projectcoverage} shows the statement coverage (``SC'') and branch coverage (``BC'') of the tests generated by \ourtool{} and all the baselines over the entirety of each project. We can find that \ourtool{} achieves comparable and even better coverage than the baselines in the project-level evaluation setting. We further inspect some cases and analyze the potential reason for the observation. In particular, compared to the deep-learning-based technique \athenatest{} and the basic \chatgpt{}, \ourtool{} is capable of generating more compilable and executable tests, thus inherently resulting in higher coverage; for the search-based technique \evosuite{}, although it is able to generate more executable tests, its generated tests fall short in containing unnatural inputs and thus cannot reach some specific statements. }

\begin{figure}[htb]
    \centering
    \includegraphics[width=0.7\textwidth]{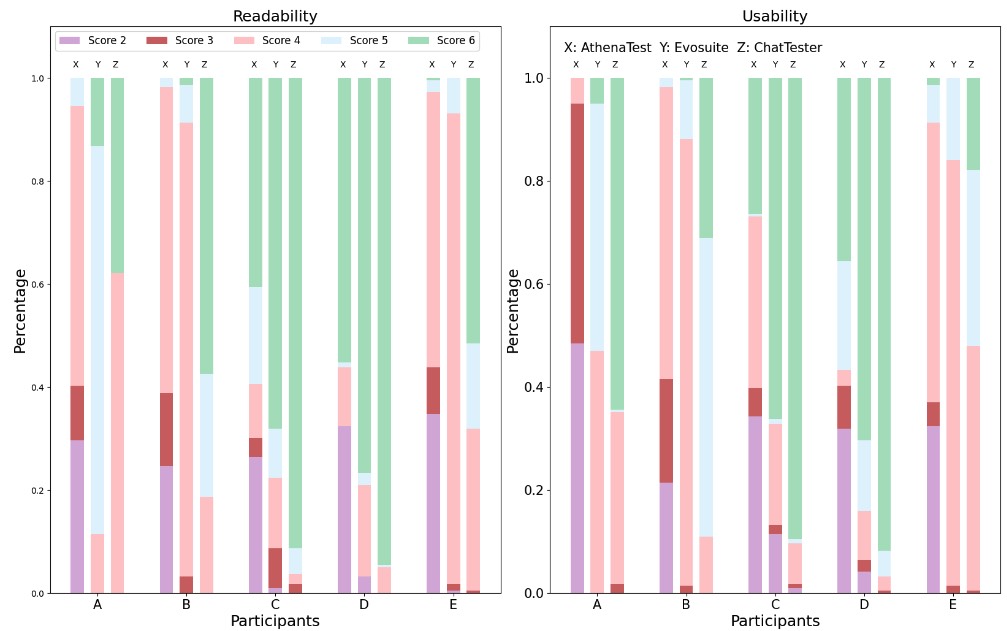}
    \caption{Score Distribution of Readability and Usability}
    \label{fig:projectUser}
\end{figure}

\parabf{\revised{Readability and Usability.}}
\revised{Figure~\ref{fig:projectUser} presents the overall score distribution of the readability and usability. In particular, A to E groups represent different participants, and X/Y/Z represent different test generation techniques. As shown in the figure, all the participants find that \ourtool{} generate more tests with better readability and usability than baselines. In addition, based on the scoring details of naming intuitiveness, code layout, assertion quality, and adoption efforts, we find that participants consider the \ourtool{}-generated tests with more intuitive naming styles and more structured code (thus with better readability), and participants also consider the \ourtool{}-generated tests with better assertions and less adoption efforts (thus with better usability).}

\begin{mdframed}[linecolor=gray,roundcorner=12pt,backgroundcolor=gray!15,linewidth=3pt,innerleftmargin=2pt, leftmargin=0cm,rightmargin=0cm,topline=false,bottomline=false,rightline = false]
\textbf{Project-level Evaluation Summary:} \ourtool{} consistently outperforms baselines in coverage, readability, and usability in the project-level evaluation setting.
\end{mdframed}

\section{Threats to Validity}
(i) One threat lies in the randomness in \chatgpt{}. To alleviate this issue, we repeat our experiments three times and present the average results when automatically evaluating the effectiveness of \ourtool{}. We do not repeat our experiments in the empirical study, due to the large manual efforts involved in the user study. However, we actually observe similar correctness results of the tests generated by \chatgpt{} on two different datasets (Table~\ref{table:rq1:correctness} and Table~\ref{table:rq4}), indicating the consistency of our results. 
(ii) Another threat is in the benchmarks used in this work. Our findings might not generalize to other datasets. To eliminate this issue, we construct our datasets to include more high-quality projects and diverse focal methods and test methods, and we evaluate the proposed \ourtool{} on a different evaluation dataset to avoid overfitting issues. \revised{In addition, to evaluate the generalization capability of \ourtool{} on different programming languages, we further perform an additional evaluation of  \ourtool{} on a Python dataset, HumanEval~\cite{HumanEval}. In particular, we leverage \ourtool{} to generate tests for 164 Python methods in HumanEval. We observed the consistent effectiveness of \ourtool{} by improving the successful execution rate of \chatgpt{}-generated tests from 31.7\% to 43.2\%. The results indicate the generalization of \ourtool{} to other programming languages.} 
(iii) Another threat is the potential data leakage of the manually-written tests being part of the training data in \chatgpt{}, which might lead to the overestimation of \chatgpt 's capability in test generation. 
\revised{In fact, we find there is quite long Levenshtein edit distance between the \ourtool{}-generated tests and manually-written tests (\ie{} average of 1,027 characters and median of 609 characters), and only 5\% of \ourtool{}-generated tests have less than 200-characters Levenshtein distance to manually-written ones, implying that directly copying from training data might be less prevalent in our benchmark.}

\section{Related Work}~\label{sec:related}
\vspace{-5mm}
\subsection{ChatGPT for SE tasks}
\chatgpt{} has attracted wide attention due to its outstanding capability of solving various tasks~\cite{fraiwan2023review, li2023evaluating, zhao2023survey, DBLP:journals/corr/abs-2302-03287}, including the SE tasks~\cite{DBLP:journals/corr/abs-2301-08653, dong2023self, chen2023gptutor,DBLP:journals/corr/abs-2307-07221}, \eg{} program repair~\cite{DBLP:journals/corr/abs-2301-08653, DBLP:journals/corr/abs-2304-00385}, code generation~\cite{dong2023self, gao2023makes, ren23}. 
Our work performs the first study to explore the \chatgpt{}'s ability for unit test generation; and then proposes \ourtool{}, a novel \chatgpt -based unit test generation approach, which improves the quality of the tests generated by \chatgpt{}.

\subsection{LLMs for Test Generation}~\label{sec:related:test}
One typical category of LLM-based test generation techniques mainly regard test generation as a neural machine translation problem~\cite{tufano2020unit, nie2023learning} (\ie{} from the focal method to the corresponding test prefix or the test assertion) and fine-tune the LLMs on the test generation dataset. For example, \athenatest{}~\cite{tufano2020unit} fine-tunes BART~\cite{chipman2010bart} on a test generation dataset where the input is the focal method with the relevant code context while the output is the complete test case. Our results show that \chatgpt{} outperforms \athenatest{} in generating tests of higher correctness and coverage.

More recently, given the rapid development of instruction-tuned LLMs, there are an increasing number of test generation techniques that leverage instructed LLMs via suitable prompts instead of fine-tuning models~\cite{DBLP:conf/issta/DengXPY023, DBLP:conf/icse/DengXYZY024, DBLP:journals/corr/abs-2308-04748}. 
\revised{For example, breaking down complex tasks into smaller tasks (such as chain of thought reasoning~\cite{wei2022chain} and tree of thought reasoning~\cite{yao2024tree}) has been demonstrated as effective prompting strategies, which are also high-level ideas inspiring \ourtool{}.} 
Nashid et al.~\cite{DBLP:conf/icse/NashidSM23} propose a retrieval-based prompt construction strategy that queries Codex with few shots to generate test assertions. Different from their work, our work focuses on the zero-shot learning scenario and generates  complete tests of both test prefixes and test assertions. CODAMOSA~\cite{lemieux2023codamosa} enhances traditional search-based techniques by using tests generated by Codex to escape from the ``plateaus'' during the search procedure. In fact, \ourtool{} is complementary to CODAMOSA, since \ourtool{} aims at generating one test for the given focal method via LLM while CODAMOSA aims at generating a set of tests for the given module via the combination of LLM and search-based algorithm. Therefore, \ourtool{} could provide higher-quality of tests as the seed for CODAMOSA during its search procedure. LIBRO~\cite{kang2022large} leverages Codex to generate tests for the given bug report. Different from LIBRO, our work focuses on the test generation scenario without bug reports.  Li et al.~\cite{ase2023Li} propose differential prompts to generate failure-inducing test inputs via \chatgpt{}, which focuses on generating \textit{test inputs} via LLMs, while our work focuses on generating a complete unit test case of both test inputs and test code (\eg{} the test prefix).
\revised{Ring System~\cite{joshi2023repair} and TestPilot~\cite{schafer2023empirical} also employ LLM for test generation and bug repair, respectively. Ring System adopts a one-turn approach focusing on a specific code snippet, whereas our tool, \ourtool{}, leverages iterative refinement through multi-round dialogues with LLMs and incorporates additional code context for enhanced accuracy. TestPilot differs from \ourtool{} in its feedback mechanism by searching for code snippets using the focal methods. In contrast, \ourtool{} extracts and utilizes additional contextual information from error messages for refining test generation.}

\section{Conclusion}
In this work, we perform the first empirical study to evaluate \chatgpt 's capability of unit test generation, by systematically investigating the correctness, sufficiency, readability, and usability of its generated tests.
We find that the tests generated by \chatgpt{} still suffer from correctness issues, including diverse compilation errors and execution failures (mostly caused by incorrect assertions); but the passing tests resemble manually-written tests by achieving comparable coverage, readability, and even sometimes developers' preference.  
Inspired by our findings above, we further propose \ourtool{}, which leverages \chatgpt{} itself to improve the quality of its generated tests.  Our evaluation demonstrates the effectiveness of \ourtool{} by generating 34.3\% more compilable tests and 18.7\% more tests with correct assertions than the default \chatgpt{}.

\balance
\bibliographystyle{ACM-Reference-Format} 

\bibliography{ref}

\end{document}